\documentclass[aps,prb,reprint,amsmath,amssymb,groupedaddress]{revtex4-2}
\usepackage{graphicx}% Include figure files
\usepackage{dcolumn}% Align table columns on decimal point
\usepackage{bm}% bold math
\usepackage{color}
\usepackage{siunitx}
\usepackage{caption}
\usepackage{subcaption}
\usepackage{hyperref}

\usepackage{todonotes}
\usepackage{threeparttable}
\usepackage{gensymb}
\usepackage{natbib}
\usepackage{url}
\usepackage{textcase}

\begin{document}

\title{Effects of the surface termination and oxygen vacancy positions and on LaNiO$_{3}$ ultra-thin films: First-principles study} 
%\hyowon{What about? Effects of the surface termination and the oxygen vacancy position on LaNiO$_{3}$ ultra-thin films: A first-principle study}

\author{Xingyu Liao$^{1}$ and Hyowon Park$^{1,2}$}
%\thanks{}
\affiliation{$^1$Department of Physics, University of Illinois at Chicago, Chicago, IL 60607, USA \\
$^2$Materials Science Division, Argonne National Laboratory, Argonne, IL, 60439, USA }


\date{\today}

\begin{abstract}
While ultra-thin layers of the LaNiO$_3$ film exhibit a remarkable metal-insulator transition as the film thickness becomes smaller than a few unit cell (u.c.),
the formation of possible oxygen vacancies and their effects on the correlated electronic structure have been rarely studied using first-principles.
Here, we investigate the effects of the surface termination and the oxygen vacancy position on the electronic properties and vacancy energetics of LaNiO$_3$ ultra-thin films under the compressive strain using density functional theory plus U (DFT+U). 
We find that oxygen vacancies can be easily formed in the Ni layers with the NiO$_2$ terminated surface (0.5 u.c. and 1.5 u.c. thickness) compared to the structures with the LaO terminated surface and the in-plane vacancy is energetically favored than the out-of-plane vacancy. When two vacancy sites are allowed, the Ni square plane geometry is energetically more stable in most cases as two oxygen vacancies tend to stay near a Ni ion.  
Strong anisotropy between the in-plane and out-of-plane  vacancy formation as well as the layer and orbital dependent electronic structure occur due to strain, surface termination, charge reconstruction, and quantum confinement effects.
The in-plane vacancy of the NiO$_2$ terminated structure is favored since the released charge due to the oxygen vacancy can be easily accommodated in the $d_{x^2-y^2}$ orbital, which is less occupied than the $d_{z^2}$ orbital. 
Remarkably, the oxygen vacancy structure containing the Ni square-plane geometry becomes an insulating state in DFT+U with a sizable band gap of 1.2eV because the large crystal field splitting between $d_{z^2}$ and $d_{x^2-y^2}$ orbitals in the square-plane favors an insulating state and the Mott insulating state is induced in other Ni sites due to strong electronic correlations.
In the thin-film structure without oxygen vacancies, the strong correlation effect in DFT+U drives a pseudo-gap ground state at the Fermi energy, similarly as the experimental photo-emission spectra, however the variation of the DFT+U electronic structure depending on the surface termination becomes weaker compared to those obtained in DFT.
\end{abstract}

%\keywords{}
\maketitle

%%%%%%%%%%%%%%%%%%%%%%%%%%% Introduction %%%%%%%%%%%%%%%%%%%%%%%%%%%%%%
\section{Introduction}
Rare earth nickelates, $R$NiO$_3$ ($R$=rare earth ions) have attracted significant research interests due to their rich electronic and magnetic properties driven by the strong electron-electron interaction of Ni 3$d$ orbitals~\cite{doi:10.1146/annurev-matsci-070115-032057,RevModPhys.70.1039}.
The rich electronic phase diagram of nickelates include the Mott transition induced by the Ni-O bond-disproportionation~\cite{PhysRevLett.109.156402,PhysRevLett.112.106404}, magnetism~\cite{PhysRevLett.115.036401}, charge ordering~\cite{PhysRevB.64.094102}, the non-Fermi liquid phase~\cite{PhysRevLett.94.226602}, the metal-to-insulator transition (MIT) induced by oxygen vacancies~\cite{MORIGA1995252_LNO2.6,Sanches_MIT}, orbital polarization~\cite{PhysRevLett.100.016404}, and emergent superconductivity~\cite{Hwang2019}.

Understanding the role of electronic correlations in the MIT occurring under various structural and chemical environments such as the thin-layer deposition, strains, and the oxygen vacancy control can open a new revenue to apply nickelates toward various functional devices including catalysts~\cite{doi:10.1021/jacs.5b11713}, field effect transistors~\cite{gunkel_oxygen_2020}, solid-oxide fuel-cells~\cite{BANNER2021230248}, and neuromorphic devices~\cite{andrews_building_2019}.  
Although bulk LaNiO$_3$ remains metallic even at very low temperatures, its electronic and magnetic properties can be dramatically changed by oxygen vacancy formations or the synthesis of thin film structures. As the oxygen vacancy concentration increases, bulk LaNiO$_{3}$ undergoes the MIT as well as the magnetic transition from paramagnetic to ferromagnetic structure
\cite{LNO_FM_AFM_2018, Sanches_MIT, Moriga_structure, MORIGA1995252_LNO2.6}.
In the form of ultra-thin films, LaNiO$_3$ shows the remarkable correlation driven MIT as the film thickness decrease since the electronic correlations are enhanced for electrons confined in the quasi-two dimensional limit~\cite{King2014,PhysRevLett.106.246403}. 
Engineering of the thin film to tune the structural and electronic properties of Ni layers can explore novel material properties of strongly correlated nickelates, which can not be obtained in bulk.
In particular, the electronic structure of the two-dimensional Ni-O square-plane geometry in $R$NiO$_2$ ($R$=Pr, Sr, and La) has been a subject of intensive research recently due to superconductivity emerged in this infinite-layer structure~\cite{Hwang2019,doi:10.1021/acs.nanolett.0c01392,doi:10.1126/sciadv.abl9927}.

Various experimental groups have investigated both transport and spectroscopic properties of LaNiO$_3$ thin films under various strains, film thickness, and surface terminations to understand the correlated electronic structure in the ultra-thin limit of Ni layers~\cite{PhysRevB.87.075132, doi.org/10.1038/nnano.2014.59nature, apldoi10.1063/1.3309713, apldoi10.1063/1.3269591, apldoi10.1063/1.5143316,Natcomm18,Qiao_2011,PhysRevB.98.014105,PhysRevApplied.2.054004}.
Sakai $et$ $al$~\cite{PhysRevB.87.075132} studied the photon emission spectrum (PES) and the resistivity of LaNiO$_3$ thin films, and showed the Ni-derived peaks decrease or disappear as the film thickness becomes less than 4 unit cell (u.c.).
Golalikhani $et$ $al$~\cite{Natcomm18} measured the resistivity and the X-ray absorption spectrum (XAS) at the oxygen $K$ edge of a-few-layer of LaNiO$_3$ ultra-thin films. They found the non-monotonic behavior of the resistivity as a function of the film thickness showing that the NiO$_2$ surface termination favors the insulating state than the LaO surface termination.
The increase of the resistivity under the low oxygen pressure implies oxygen vacancies also plays a role in resulting in insulating nature of LaNiO$_3$ ultra-thin films.

Qiao $et$ $al$~\cite{Qiao_2011} observed and distinguished Ni$^{3+}$ and Ni$^{2+}$ states via X-ray photoelectron spectroscopy deconvolution analysis in correlated LaNiO$_{3-x}$ films on Si substrate.
Their work showed that the LaNiO$_{3-x}$ film has higher resistivity with the higher oxygen vacancy level and the lower Ni$^{3+}$/Ni$^{2+}$ ratio. 
Anada $et$ $al$~\cite{PhysRevB.98.014105} argued that the nature of the insulating state in the LaNiO$_3$ thin film on the LaAlO$_3$ substrate is the covalent insulator driven by the covalent Ni-O bond length measured by the surface X-ray diffraction.
LaNiO$_3$ thin films also shows the dependence of surface terminations on transport properties~\cite{PhysRevApplied.2.054004}.

Although many experimental evidences show that the correlated electronic structure of LaNiO$_{3}$ ultra-thin films can be strongly affected by structural modifications of different surface terminations and the film thickness with or without oxygen vacancies,
theoretical studies of such effects in thin films by adopting first-principles of realistic slab structures have been lacking.
Previous theoretical studies used a model Hamiltonian with anisotropic hopping parameters to treat the thin film effect Ni layers using dynamical mean field theory (DMFT)~\cite{PhysRevB.87.075132} and a slave-roter Hartree-Fock method~\cite{PhysRevLett.110.126404}.
Surface termination effect of LaNiO$_3$ thin films in a realistic slab structure has been studied using density functional theory (DFT)~\cite{PhysRevApplied.2.054004}, DMFT~\cite{Natcomm18}, and  DFT+U~\cite{doi:10.7566/JPSJ.87.114704}.
Previous first-principles studies of the oxygen vacancy effect on LaNiO$_3$ in both bulk~\cite{xingyuliao} and thin film~\cite{PhysRevB.102.165411} structures showed that the correlated electronic structure and the MIT are strongly affected by the vacancy effect. 

In this work, we did a systematic study of the effects of oxygen vacancy positions and the surface termination on the correlated electronic properties and the charge reconstruction of LaNiO$_3$ ultra-thin films under compressive strains using first-principles by constructing the slab crystal structure to include realistic surface and substrate effects. 
We find that the electronic structure of LaNiO$_{3}$ ultra-thin layers without explicit oxygen vacancies can exhibit the strong modification from the bulk results depending on the surface termination and the film thickness.
In particular, the Ni $d_{z^2}$ orbital is mostly affected as its band width becomes narrower and it is more occupied than the Ni $d_{x^2-y^2}$ orbital due to the strain and strong quantum confinement effects.  
The charge reconstruction occurs mostly at the top surface of LaNiO$_3$ thin films due to the polar nature of layers.
The NiO$_2$ terminated surface layer has the charge state of [NiO$_2$]$^{0.63-}$, which has 0.32e less charge compared to the bulk case while the LaO terminated surface layer has the charge state of [LaO]$^{0.7+}$, which 0.25e more charge compared to bulk.
This charge transfer mostly affects the oxygen $p$ DOS as the oxygen spectra moves close to the Fermi energy in the NiO$_2$ terminated structure.
The in-plane oxygen vacancy is energetically favored in the NiO$_2$ terminated structure since the released charge from the vacancy can be easily accommodated in this structure compared to the LaO terminated case.
We find that the oxygen vacancy induced MIT in the thin film can be strongly affected by the vacancy position since the formation of the Ni-O square-planar structure favors the insulating phase compared to other geometries induced by vacancies.

%%%%%%%%%%%%%%%%%%%%%%%%%%% Computational method %%%%%%%%%%%%%%%%%%%%%%%%%%% 
\section{\label{sec:method}Computational Method}

We studied the structural property, the correlated electronic structure, and the vacancy energetics of LaNiO$_3$ thin films by adopting both DFT and DFT+U implemented within 
Vienna Ab-initio Simulation Package (VASP)~\cite{vasp1, vasp2}. 
The Perdew-Burke-Ernzerhof (PBE) Sol functional~\cite{PBEsol} has been used to treat the exchange and correlation energy.

\begin{figure}[h]
    \begin{subfigure}{1.\linewidth}
       \centering
       \includegraphics[width=1.\linewidth]{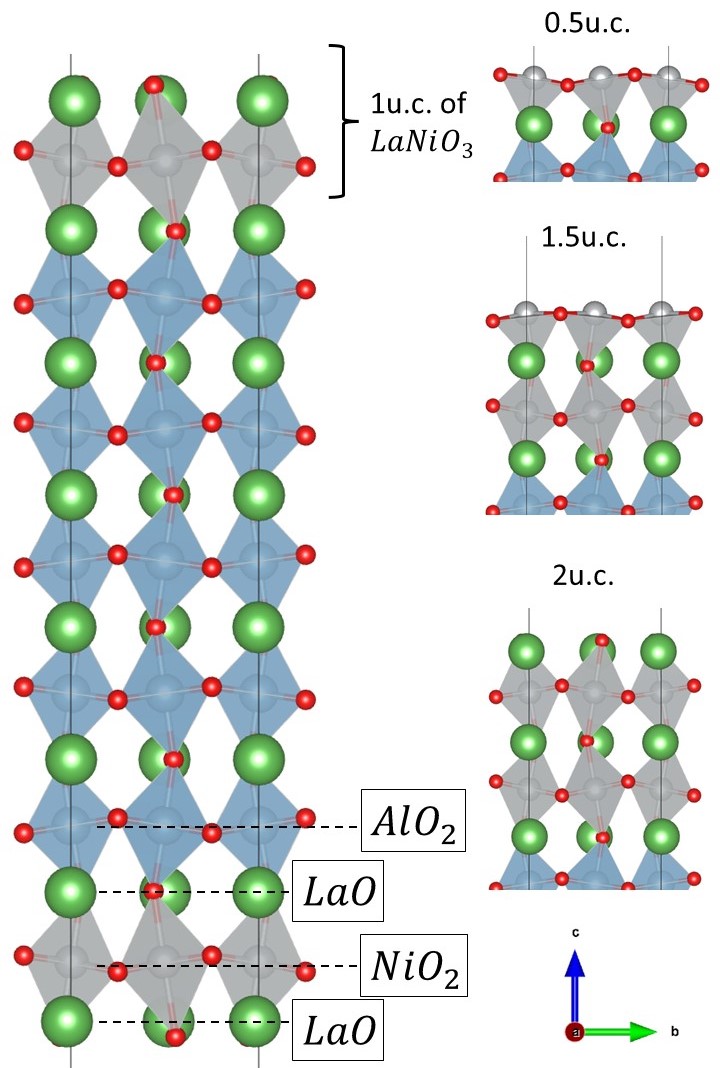}
    \end{subfigure}
    \begin{subfigure}{1.\linewidth}
       \centering
       \includegraphics[width=1.\linewidth]{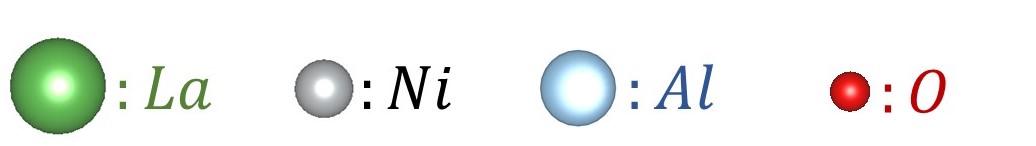}
    \end{subfigure}
\caption{Left: Relaxed crystal structure of the 1.0 u.c. LaNiO$_3$ layer on the LaAlO$_3$ substrate. The mirror symmetry about the $a-b$ plane has been imposed to the simulation of the slab structure to avoid the necessity of building a semi-infinite substrate. 
Right: Examples of the different LaNiO$_3$ thickness with the NiO$_2$ (0.5 and 1.5 u.c.) or LaO (1.0 and 2.0 u.c.) surface terminations considered in this paper.}
\label{fig:struct}
\end{figure}

To simulate realistic thin-film effects, we build a slab structure including the LaNiO$_3$ thin layers, the LaAlO$_3$ substrate, and the vacuum (see Fig.$\:$\ref{fig:struct}).
We also impose the mirror symmetry about the $a-b$ plane to avoid the polar nature of the substrate at the bottom interfaced to the vacuum and to reproduce the semi-infinite nature of the substrate.
The slab unit cell contains two Ni (Al) ions per layer to accommodate octahedral rotations.
The vacuum length is at least 15\AA$\:$ for all structures considered here.
For crystal structures containing oxygen vacancies, we consider either one vacancy site per LaNiO$_3$ thin layers (see Fig.$\:$\ref{fig:Ovac}) or two vacancy sites per LaNiO$_3$ thin layers in a supercell with four Ni ions per layer (see Fig.$\:$\ref{fig:Ovac_sc}).
To study structural properties, we relax the internal ionic positions with the convergence criterion of the atomic force smaller than 0.01 eV/\AA$\:$ while fixing the shape and the volume of the slab.
Here, we relax ionic positions of the LaNiO$_3$ thin layers and the first interfacial layer of LaAlO$_3$ while other ionic positions are fixed.
Due to the large size of the unit cell, a 4$\times$4$\times$1 $k-$point grid has been used for the supercell containing two vacancy sites and 8$\times$8$\times$1  $k-$point grid has been used for the slab and the unit cell containing one vacancy site. 

To include the strong correlation effects of Ni ions, we adopt DFT+U for structural relaxations and the electronic structure calculations.
To treat the local Coulomb interaction in DFT+U, we adopt the simplified rotationally invariant interaction formula introduced by Dudarev $et$ $al$~\cite{PhysRevB.57.1505}. 
In this formula, the local interaction $U_{eff}$ is defined effectively by U-J, namely the difference between the local Hubbard interaction U and the Hund's coupling J. 
We used $U_{eff}$=6eV for all DFT+U calculations since this value is consistent with the previous self-consistent constrained DFT result~\cite{PhysRevB.84.144101}. 
In DFT+U calculations, we impose magnetism in Ni ions with the ferromagnetic (FM) order since we find that the FM energy is always lower than the antiferromagnetic (AFM) energy in all thin films (see Table.\ref{tbl:Ttl_E}).
Our finding is also consistent with the experimental measurement of bulk LaNiO$_{3-x}$ with multiple vacancy concentration $x$ values showing the FM order~\cite{LNO_FM_AFM_2018, Sanches_MIT, Moriga_structure, MORIGA1995252_LNO2.6} except near $x=0.5$.
For the electronic structure calculation, we  also compare both DFT and DFT+U results to reveal the strong correlation effects of Ni ions.
To study the charge reconstruction effect, we also perform the Bader charge analysis of integrating the charge density of valance electrons within a layer~\cite{doi:10.1063/1.2161193, HENKELMAN2006354}.

\begin{table}[h]
    \centering
    \caption{\label{tbl:Ttl_E} Total energy (per one Ni) difference between AFM and FM ($E_{AFM}-E_{FM}$) of LaNiO$_3$ thin film structures obtained using DFT+U.}
    \begin{tabular}{p{0.36\linewidth}p{0.14\linewidth}p{0.14\linewidth}p{0.14\linewidth}p{0.14\linewidth}}
    \hline\hline
                        &   0.5u.c. &1.0u.c.    &1.5u.c.    &2.0u.c.\\
    \hline
    $E_{AFM}-E_{FM}$(eV)&   0.16	&0.07		&0.04		&0.12\\
    \hline
    \end{tabular}
\end{table}

%%%%%%%%%%%%%%%%%%%%%%%%%%%% Results %%%%%%%%%%%%%%%%%%%%%%%%%%%% 
\section{Results}
\subsection{LaNiO$_3$ thin film: Structural information \label{subsec:rlx_struct}}

\begin{table*}[t]
    \centering
    \caption{\label{tbl:struct}In-plane (IP) and out-of-plane (OP) Ni-O bond lengths $d$ and Ni-O-Ni angle $\theta$ in 0.5-2.0 u.c. thin films obtained using the DFT+U relaxation calculation. Calculated structural parameters are compared with experimental parameters in the 95\AA$\:$ film obtained from Ref.\cite{PhysRevB.82.014110}. }
    \begin{tabular}{p{0.16\linewidth}p{0.03\linewidth}p{0.13\linewidth}p{0.13\linewidth}p{0.13\linewidth}p{0.13\linewidth}p{0.13\linewidth}p{0.08\linewidth}}
    \hline\hline
       &&Exp (95\AA) &0.5u.c. & 1.0u.c.&1.5u.c.&2.0u.c. & Bulk\\ [1.0ex] 
    \hline
       $d_{IP}$   (\AA)         && 1.916$\pm$0.005  &1.927  &1.918  &1.925      &1.936  &1.968\\ 
       $d_{OP}$ (\AA)       && 1.949$\pm$0.002  &2.028  &2.067  &1.975      &2.125  &1.968\\ 
       $\theta_{IP}$ ($^\circ$)     && 164.0$\pm$2.0    &162.7  &163.0  &163.5      &162.4  &160.1\\
       $\theta_{OP}$($^\circ$)  && 175.2$\pm$0.6    &N/A    &N/A    &163.0      &161.5  &160.1\\
     \hline\hline
    \end{tabular}
\end{table*} 

While bulk LaNiO$_3$ has a rhombohedral crystal structure with the $a^-a^-a^-$ rotation pattern of the Ni octahedra using the Glazer notation~\cite{Glazer_notation}, 
thin film structures show the distinct rotation patterns depending on strain effects under different substrates.
LaNiO$_3$ films on the compressive LaAlO$_3$ substrate shows the $a^-a^-c^-$ rotation pattern as confirmed by the X-ray diffraction measurement~\cite{PhysRevB.82.014110}.
This octahedral rotation is accompanied by the elongation of the apical Ni-O bond length and the enhanced out-of-plane Ni-O-Ni rotation angle.
Here, we built slab structures with different thickness (0.5-2.0 u.c.) of LaNiO$_3$ thin films on the LaAlO$_3$ substrate. 
As shown in Fig.$\:$\ref{fig:struct}, both 1.0 u.c and 2.0 u.c structures have the LaO capping layer while both 0.5 u.c. and 1.5 u.c. structures have the NiO$_2$ layer at the top surface, thus different surface termination effects can be compared.
All structures obey the $a^-a^-c^-$ rotation pattern as the experiment.

To simulate both the surface and substrate effects of LaNiO$_3$ thin films, we build a slab structure including the vacuum, LaNiO$_3$ layers, and LaAlO$_3$ layers.
First, we relax the LaAlO$_3$ bulk structure using DFT and build a $\sqrt{2}\times\sqrt{2}\times5$ pseudo-cubic supercell to accommodate the $a^-a^-c^-$ octahedral rotation and the substrate effect along the $c$ direction.
Then, we relax the LaNiO$_3$ bulk structure using DFT+U. We find that the in-plane lattice constant of LaNiO$_3$ shows the -1.5$\%$ compression compared to that of LaAlO$_3$ confirming that the LaAlO$_3$ substrate produces the compressive strain on the LaNiO$_3$ thin films.
Our calculated -1.5$\%$ strain effect is also close to a previous experimental measurement of the -1.1$\%$ compressive strain of the LaAlO$_3$ substrate~\cite{PhysRevB.82.014110}. 
Next, LaNiO$_3$ layers with the 0.5-2.0 unit cell (u.c.) thickness are placed on the top of the LaAlO$_3$ substrate.
LaNiO$_3$ layers exhibit the same $a^-a^-c^-$ octahedral rotation as LaAlO$_3$.
Since the structure of the LaAlO$_3$ substrate will be similar as the bulk structure, we relax only the LaNiO$_3$ layers and the first layer of LaAlO$_3$ at the interface while the remaining LaAlO$_3$ layers are kept static.
Moreover, the mirror symmetry about the $a$-$b$ plane was imposed to simulate the semi-infinite substrate effect and avoid the artificial polar nature at the surface of the substrate.
To simulate the vacuum effect along the $c$-axis, we build the slab structure with a large lattice constant of 45-60\AA$\:$ along the $c-$direction ensuring that different thin film structures will have the vacuum length of at least 15\AA$\:$.

In Table~\ref{tbl:struct}, we compare the relaxed structural parameters of the slab structure with the experimentally measured values. %, as shown in Table:\ref{tbl:struct}.
Since there are multiple Ni-O bond lengths and Ni-O-Ni bond angles in structures from 0.5 u.c. to 2.0 u.c., we provide average values for these structures. Experimental values were measured from 95\AA-thick film while our films are thinner than 8\AA. 
Our calculated in-plane (IP) bond lengths are 
relatively closed to experimental measurements while the out-of-plane (OP) bond lengths are more elongated in calculations. 
This is because our calculated bond lengths in bulk LaNiO$_3$ (1.968\AA) are already larger than experiments (1.935\AA)~\cite{PhysRevB.82.014110} and the crystal volume is overestimated in calculations. With the compressive strain, the OP Ni-O bond length becomes longer than the IP one to conserve the crystal volume and the ratios of $d_{OP}/d_{IP}$ are $\sim 1.03-1.09$. 
The calculated OP Ni-O-Ni bond angles are also reduced from experiments to accommodate the elongated OP bond length. 
Due to the polar nature of the surface, one can also expect the Ni-O buckling distortion on the surface.
We find that our Ni-O octahedra are slightly tilted and rotated along the $c$ axis, therefore the buckling distortion is much smaller than the previously reported value calculated without octahedral tilting and rotation~\cite{PhysRevApplied.2.054004}.

\subsection{LaNiO$_3$ thin film: Electronic structure and charge reconstruction}\label{subsec:ES_film}

\begin{figure}
       \centering
       \includegraphics[width=1\linewidth]{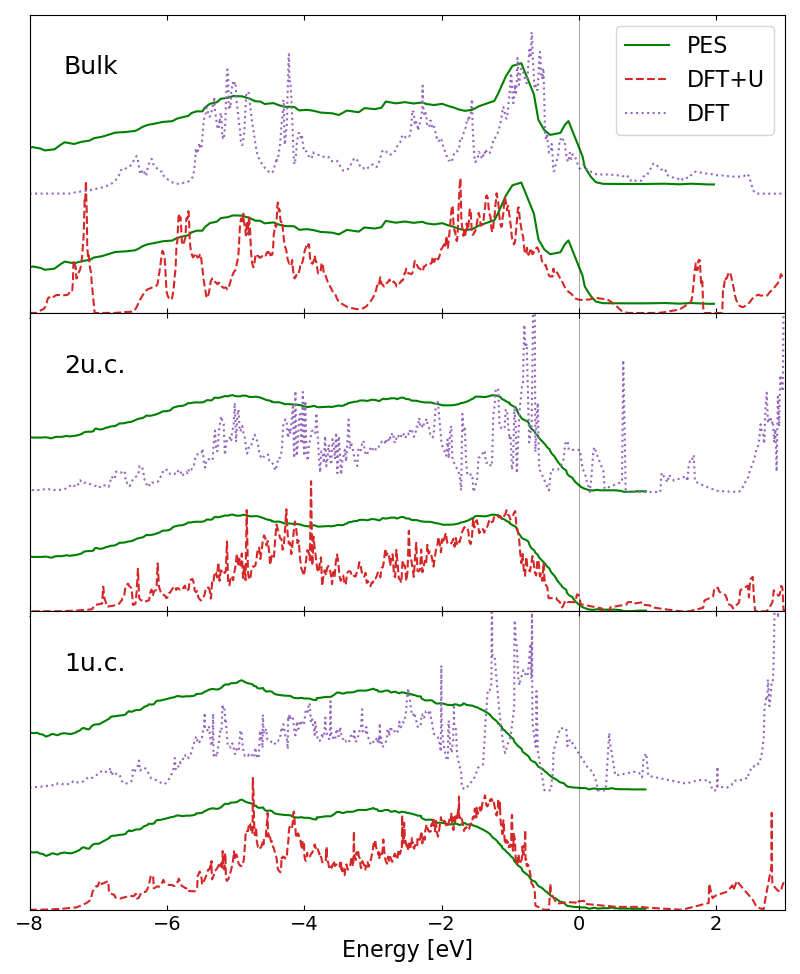}
\caption{Total density of states (DOS) for LaNiO$_3$ bulk (top panel) and thin films with the 2.0 u.c. (middle panel) and 1.0 u.c. (bottom panel) thickness calculated using DFT and DFT+U. Calculated spectra are compared with the PES measurement, which is extracted from Ref.$\:$\onlinecite{LaNiO3_PES_XAS, PhysRevB.87.075132}.} 
\label{fig:DOS_PES}
\end{figure}

Here, we study the evolution of electronic structure in LaNiO$_3$ films as a function of the film thickness. 
In Fig.\ref{fig:DOS_PES}, we compare the total DOS computed using DFT and DFT+U with the experimental spectra obtained from the previous PES measurements~\cite{LaNiO3_PES_XAS, PhysRevB.87.075132} for bulk, 1.0 u.c., and 2.0 u.c. of LaNiO$_3$ ultra-thin films. 
The DFT DOS is calculated with the paramagnetic order while the FM order (the lowest energy order) is imposed in DFT+U.
In bulk LaNiO$_3$, two peaks located near -0.3eV (Ni $e_g$) and -0.8eV (Ni $t_{2g}$) below the Fermi level and two broader peaks (O 2$p$ states) near -2eV and -5eV below the Fermi energy emerge.
While the overall peak positions of the PES spectra are consistent with the DFT calculation, DFT+U DOS peaks are shifted below the Fermi energy and the DOS at the Fermi level is further reduced due to the overestimation of electronic correlation effects.
The DFT+DMFT DOS in bulk LaNiO$_3$ can capture both the spectral peak positions and the $e_g$ band renormalization due to electronic correlations, as discussed in our previous work~\cite{xingyuliao}.

Now, we turn to the ultra-thin film case of both 1.0 and 2.0 u.c. layers. 
In the experimental PES spectra, the $e_g$ peak near the Fermi energy is strongly reduced and the spectral gap opens showing the MIT as the film thickness is reduced below the 2.0 u.c..
The O 2$p$ peak position is almost the same as the bulk one.
As one can see in Fig.\ref{fig:DOS_PES}, the DFT+U spectra have better fit to the experiments in the sense that the strong reduction of the $e_g$ state near Fermi level and the broader redistribution of the $t_{2g}$ state are reproduced better than the DFT spectra. The significant decrease of the DOS at the Fermi level in DFT+U is consistent with the higher resistivity measured in the ultra-thin layer of LaNiO$_3$ films compared to the bulk LaNiO$_3$ case. 
Our total DOS calculations in DFT+U show the importance of strong correlation effects to describe electronic structure of ultra-thin layers in LaNiO$_3$ films.

\begin{figure}
       \centering
       \includegraphics[width=1\linewidth]{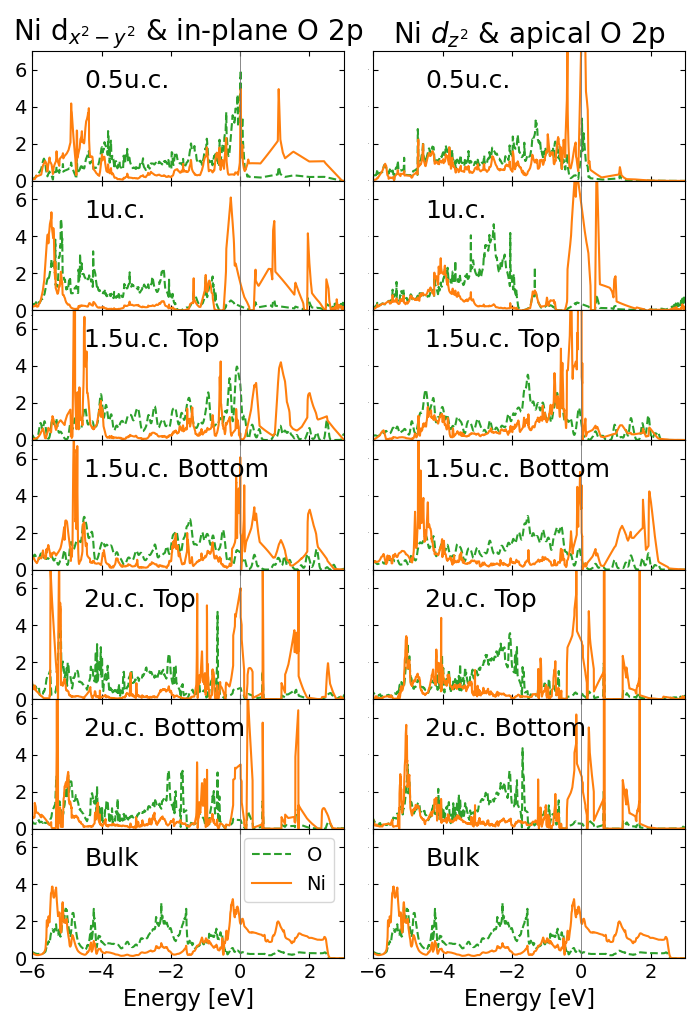}
    \caption{The orbital and layer resolved DOS computed using DFT for bulk (the bottom panel) and thin films with the 0.5-2.0 u.c. thickness. Left panel: Ni $d_{x^2-y^2}$ (orange solid line) and IP O $p$ (green dashed line) orbitals are shown. Right panel: Ni $d_{z^2}$ (orange solid line) and OP O $p$ (green dashed line) orbitals are shown.} 
\label{fig:Ni_Niall_dft}
\end{figure}

To investigate the effect of the layer thickness on the orbital-dependent physics of thin films, we plot the DOS for two Ni $e_g$ and in-plane/apical O 2$p$ orbitals of the 0.5-2.0 u.c. thin films. 
First, we plot the paramagnetic DFT DOS in Fig.\ref{fig:Ni_Niall_dft} to study the DOS evolution without treating strong correlation physics.
In the bulk DOS (see Fig.\ref{fig:Ni_Niall_dft} bottom panel), $d_{z^2}$ and $d_{x^2-y^2}$ orbitals are degenerate as Ni ions are surrounded by six O ions forming a perfect octahedron and the $e_g$ orbital symmetry is conserved under the cubic symmetry operation.
In the layer structure of thin films, this cubic symmetry is broken due to 1) the Jahn-Teller distortion of the octahedra resulting from the difference between the in-plane and out-of-plane Ni-O bond-lengths, and 2) the quantum confinement effect originating from the wide-gap insulating substrate and the vacuum. As a result, the DOS between $d_{z^2}$ and $d_{x^2-y^2}$ orbitals becomes significantly different resulting the orbital polarization effect.
Under the compressive strain, the out-of-plane Ni-O bond is longer than the in-plane one and the $d_{z^2}$ orbital energy becomes lower than the $d_{x^2-y^2}$ one.
Under the quantum confinement effect, the $d_{z^2}$ orbital reduces its kinetic energy and its band-width becomes narrower.
In the 1 u.c. case, the $d_{z^2}$ orbital is less unoccupied and shows the narrower band-width than the the $d_{x^2-y^2}$ one due to the compressive strain and the strong quantum confinement effects.
In the 2 u.c. case, this difference of the DOS between the $d_{z^2}$ and the $d_{x^2-y^2}$ orbitals becomes smaller as the quantum confinement effect becomes weaker than the 1 u.c. case.
Overall, as the film thickness decreases, bands (in particular the $d_{z^2}$ bands) near the Fermi energy become much sharper and narrower than the bulk case due to the quantum confinement effect and the reduced Ni-O hybridization of thin layers.

In bulk LaNiO$_3$, both Ni $d_{x^2-y^2}$ and $d_{z^2}$ spectra show bonding peaks near -5.7eV below the Fermi energy.
In the 2u.c. systems, the bonding peaks of Ni $d_{x^2-y^2}$ and $d_{z^2}$ orbitals show the similar position as the bulk case and both the top and the bottom spectra look similar.
As the film thickness reduces below 1.0 u.c., the $d_{z^2}$ orbital bonding peak near -5.7eV becomes much weaker and the $d_{z^2}$ band near the Fermi energy becomes strongly flat.
This shows the strongly localized nature of $d_{z^2}$ orbital in both 0.5 and 1.0 u.c..
In both 0.5 and 1.5 u.c. cases, the bonding peaks near -5.7eV are shifted to higher energy (-5.3eV). This can be attributed to the rigid band shift due to the charge redistribution depending on Ni layers and the surface termination in the LaNiO$_3$ thin film. 

To study the surface termination effect, we compare the 0.5 u.c. and the 1.0 u.c. cases. 
In the 0.5 u.c. system, the surface layer is a negatively charged NiO$_2$ layer leading to the surface polarization along with the Ni-O buckling effect. The interface between the polar and non-polar layers can lead to the divergence of the electric potential (so called the polar catastrophe scenario), and an effective charge transfer occurs to avoid this problem. A famous example is the charge transfer of 0.5$e$ occurring at the interface of LaAlO$_3$/SrTiO$_3$~\cite{Ohtomo2004}. In our case, the negatively charged NiO$_2$ surface in 0.5 u.c. can have an effective charge depletion of 0.5$e$ per Ni while an effective charge accumulation can occur on the LaO surface in 1.0 u.c..
This charge transfer is reflected in the layer-dependent DOS showing the higher bonding peak of the Ni spectra near -5.3eV and the more O $p$ states near the Fermi energy in both the 0.5 u.c. and the 1.5 u.c. top layer cases due to the effective charge depletion.
In contrast, the O $p$ states are located further below from the Fermi energy in the 1 u.c. case due to the effective charge accumulation.
The charge reconstruction also occurs between top (surface) and bottom layers in 1.5 u.c. as the $d_{z^2}$ orbital on the top layer is more occupied than the one on the bottom layer.

This sensitive change of the DOS depending on the film thickness and the surface termination implies the importance of the local chemical environment and the quantum confinement effect in thin films.
In the bulk and 2u.c. systems, all Ni ions are six-coordinated and form octahedra with surrounded O ions, as a result, $d_{z^2}$ and $d_{x^2-y^2}$ orbitals are almost degenerated. 
As the film thickness decreases, the strong quantum confinement effect suppresses the kinetic energy of Ni $d_{z^2}$ orbital significantly and the band-width of the $d_{z^2}$ band becomes much narrower and the orbital becomes more localized.
As a result, the in-plane and the out-of-plane spectra become noticeably different in the ultra-thin limit of films.
Moreover, due to the absence of apical O ions on the surface of NiO$_2$ layers, Ni ion is five-coordinated and form a pyramid structure with the neighboring O ions in both 0.5 u.c. and the 1.5 u.c. top layers. 
These layers show the noticeably different spectra with the shift of quasi-particle peak positions due to the change of local chemical environment.

\begin{figure}
       \centering
       \includegraphics[width=1\linewidth]{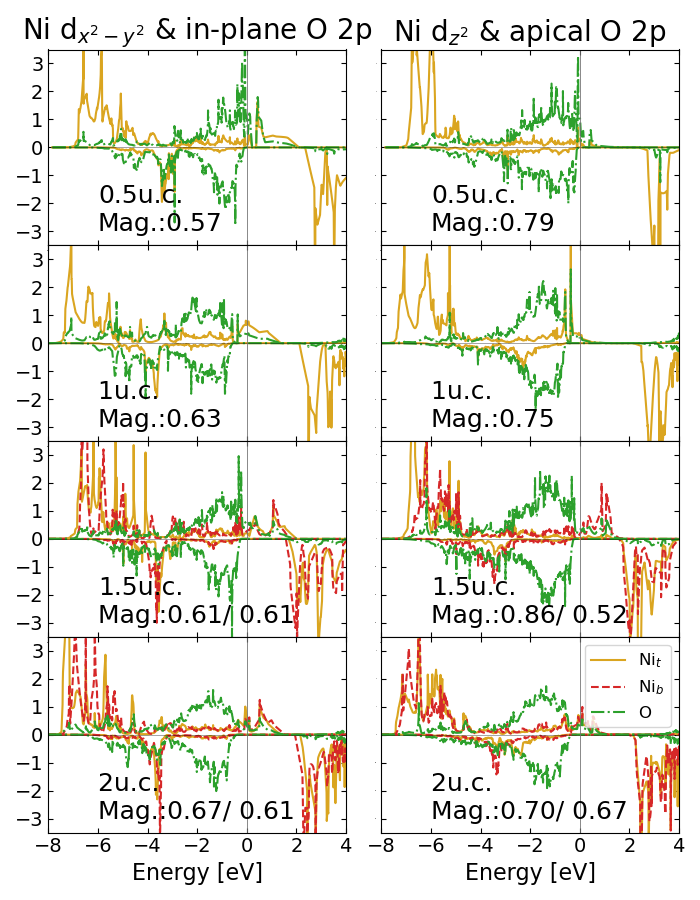}
\caption{
The orbital and layer resolved DOS computed using DFT+U for thin films with the 0.5-2.0 u.c. thickness. Left panel: top layer Ni (Ni$_t$) $d_{x^2-y^2}$ (orange solid line), bottom layer Ni (Ni$_b$) $d_{x^2-y^2}$ (red dashed line), and IP O $p$ (green dashed line) orbitals are shown. Right panel: top layer Ni (Ni$_t$) $d_{z^2}$ (orange solid line), bottom layer Ni (Ni$_b$) $d_{z^2}$ (red dashed line), and OP O $p$ (green dashed line) orbitals are shown.
Magnetic moment of each Ni ion is also displayed.}
\label{fig:Ni_dos_u}
\end{figure}

To investigate the strong correlation effect onto electronic structures of Ni ions, we plot the Ni $e_g$ and O 2$p$ DOS obtained using DFT+U in Fig.\ref{fig:Ni_dos_u}.
The obtained magnetic moment in each Ni orbital can dictate the degree of correlations at different layers.
The Ni $d_{z^2}$ orbitals show larger magnetic moments than the $d_{x^2-y^2}$ orbitals since the quantum confinement effect narrows the band-width of $d_{z^2}$ orbitals in DFT and the top Ni layer is slightly more correlated than the bottom layer.
This result is consistent with the previous DFT+DMFT result exhibiting the layer and orbital dependent correlations in the ultra-thin LaNiO$_3$ film~\cite{Natcomm18}.
In DFT+U DOS, the unoccupied part of Ni $e_g$ spectra are pushed to higher energy levels compared to the DFT DOS while the occupied part stay at the similar energy level as the DFT one. 
Overall, orbital polarization, the quantum confinement, and the layer dependence effects of the electronic structure are weaker in DFT+U compared to those in DFT.
This is also consistent with the previous study showing that orbital polarization is reduced under strong correlations of the Hund's coupling in both the bulk~\cite{PhysRevB.90.045128} and the heterostructure~\cite{PhysRevB.93.235109} forms of nickelates.

\begin{table}[h]
    \centering
    \caption{\label{tbl:BaderCHG_film}Net charge per formula unit of each layer at bulk and 0.5-2.0 u.c. thin films calculated using the Bader charge analysis.}
    \begin{tabular}{p{0.22\linewidth}p{0.15\linewidth}p{0.15\linewidth}p{0.15\linewidth}p{0.15\linewidth}p{0.1\linewidth}}
    \hline\hline
                &2u.c.  &1.5u.c.    &1u.c.  &0.5u.c.    &bulk\\ [1.0ex] 
    \hline
    LaO         &+0.72  &           &       &           &+0.95\\
    NiO$_2$     &-1.05  &-0.67      &       &           &-0.95\\
    LaO         &+0.85  &+1.01      &+0.70  &           &+0.95\\
    NiO$_2$     &-0.98  &-0.85      &-1.12  &-0.60      &-0.95\\
    \hline\hline
    \end{tabular}
\end{table} 

As discussed above, the charge reconstruction can occur on the polar surface and affect the electronic structure. To quantify this effect, we applied DFT+U in the VASP calculations and did the Bader charge analysis.
In Table.\ref{tbl:BaderCHG_film}, we list the net charge per formula unit of each film thickness from the top to the bottom Ni layers. 
In the simple ionic limit, each layer of [LaO]$^{1+}$ ([NiO$_2$]$^{1-}$) in bulk should have the net charge of +1 (-1). In the Bader charge analysis, LaO (NiO$_2$) layer in bulk has the net charge of 0.95 (-0.95), whose absolute value is slightly reduced from the ionic limit.  
In the polar catastrophe scenario, the LaO (NiO$_2$) surface should have +0.5$e$ (-0.5$e$) charge per formula unit to avoid the divergence of the electric potential. 
In our calculations, the LaO surface has the net average charge of [LaO]$^{0.71+}$ while the NiO$_2$ surface has the net average charge of [NiO$_2$]$^{0.63-}$.
This charge transfer on the surface is somewhat smaller than expected from the polar catastrophe scenario.
Nevertheless, the LaO surface is ``less positive'' (with 0.24$e$ more charge) and the NiO$_2$ surface is ``less negtive'' (with 0.32$e$ less charge) compared to the bulk case.
Remaining layers are also slightly changed from the expected bulk values.
Although the notable charge transfer occurs depending on the Ni layer and the surface termination, the Ni $d$ orbital occupancy calculated within a sphere remains close to 8.0 in all cases.
This shows that the charge redistribution mainly happens in the O $p$ orbitals.

\subsection{\label{subsec:oxygvac}Possible oxygen vacancy configurations in LaNiO$_3$ thin films}
\begin{figure}[h]
       \centering
       \includegraphics[width=1.\linewidth]{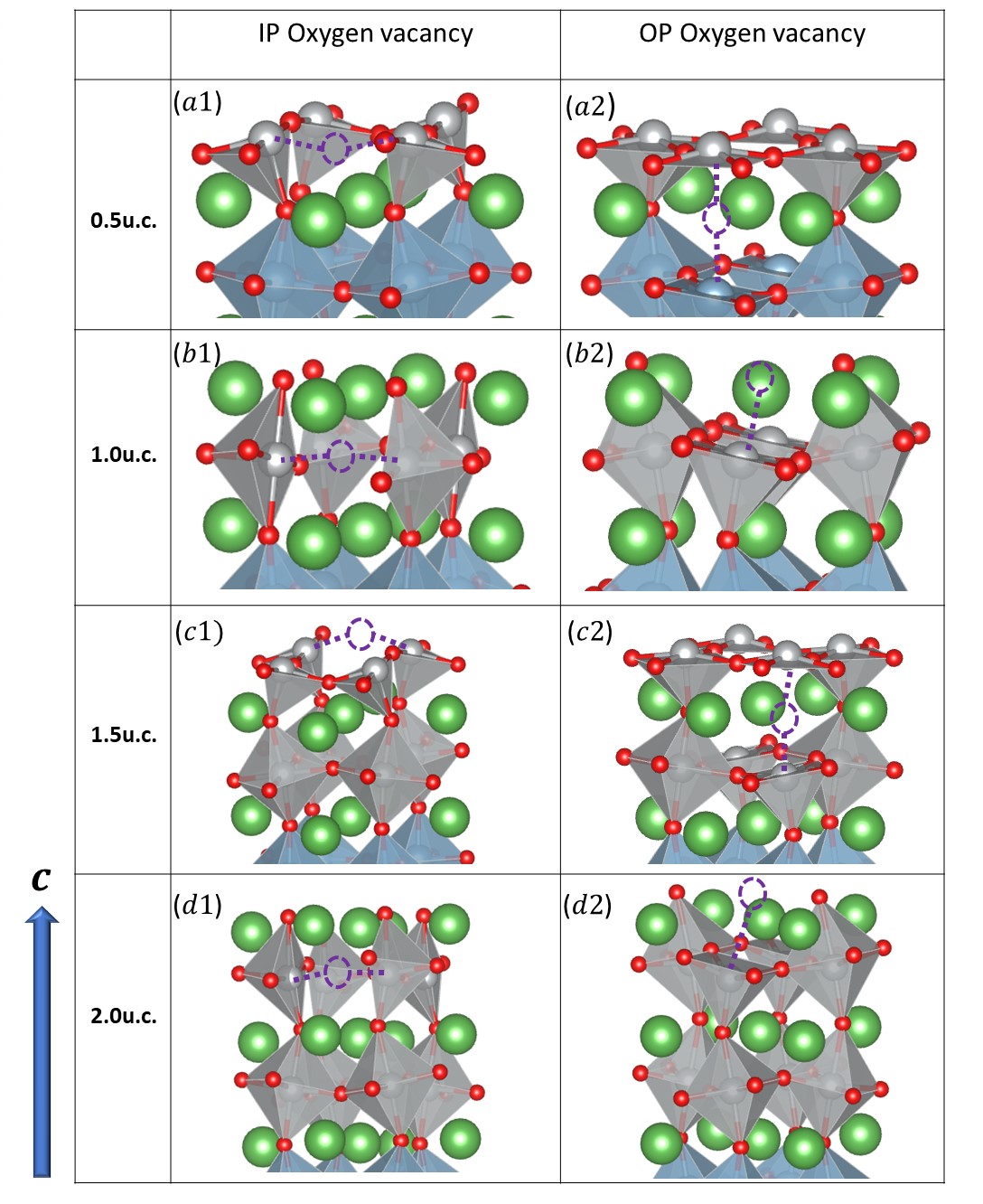}
\caption{Crystal structure of oxygen defective LaNiO$_3$ thin films with the 0.5-2.0 u.c. thickness. Dashed purple circles represent oxygen vacant sites. 
Only one oxygen-vacant site is considered for each crystal structure, possibly at IP ($[a-d]1$) or OP ($[a-d]2$) sites.}
\label{fig:Ovac}
\end{figure}

Oxygen vacancies in the perovskite structure of transition metal oxides play an important role in modifying electronic and structural properties.
They can distort the Ni octahedra and generate other types of Ni chemical environments including pyramids, tetrahedrons, and square-planes, resulting in different orbital energy splittings.
In addition to the structural modification, oxygen vacancies can donate more electrons to nearest ions along with the charge redistribution.
Thus it is important to consider oxygen-vacant sites explicitly in the calculation to understand the electronic and magnetic properties of such systems.

In bulk LaNiO$_3$, previous DFT calculation showed that two oxygen vacancies tend to stay near one Ni ion to form a NiO$_2$ square plane~\cite{DFT_study}. 
However, in the ultra-thin film case, we find that the vacancy formation energy depends sensitively on the vacancy position and the surface termination effect.
To determine the possible positions of ordered oxygen-vacant sites, we created different ultra-thin film structures with oxygen vacancies and studied the corresponding formation energies. 
First, we consider one vacant site per each unit-cell structure. 
We construct a defect-free structure having two Ni ions per layer and remove one O ion either in the LaO layer or the NiO$_2$ layer for two possible vacancy configurations. Removing the O ion in the the LaO layer creates the apical oxygen vacancy while the in-plane oxygen vacancy is obtained by removing the O ion in the  NiO$_2$ layer (see Fig.\ref{fig:Ovac}). For the 0.5 u.c. structure, 
the vacancy at the NiO$_2$ layer creates the Ni-O tetrahedron shape (see Fig.\ref{fig:Ovac}a1) while the one at LaO layer forms the Ni-O square plane (see Fig.\ref{fig:Ovac}a2). 
For the 1.5 u.c. case, we also construct  similar configurations which has IP oxygen vacancy (see Fig.\ref{fig:Ovac}c1) or OP oxygen vacancy (see Fig.\ref{fig:Ovac}c2). 
For 1 u.c. and 2 u.c. cases, the oxygen-vacant site can be on the topmost La-O or NiO$_2$ layer creating the pyramid Ni-O shape in either case (see Fig.\ref{fig:Ovac}b1,b2,d1, and d2). 
\begin{figure}[t]
       \centering
       \includegraphics[width=1.\linewidth]{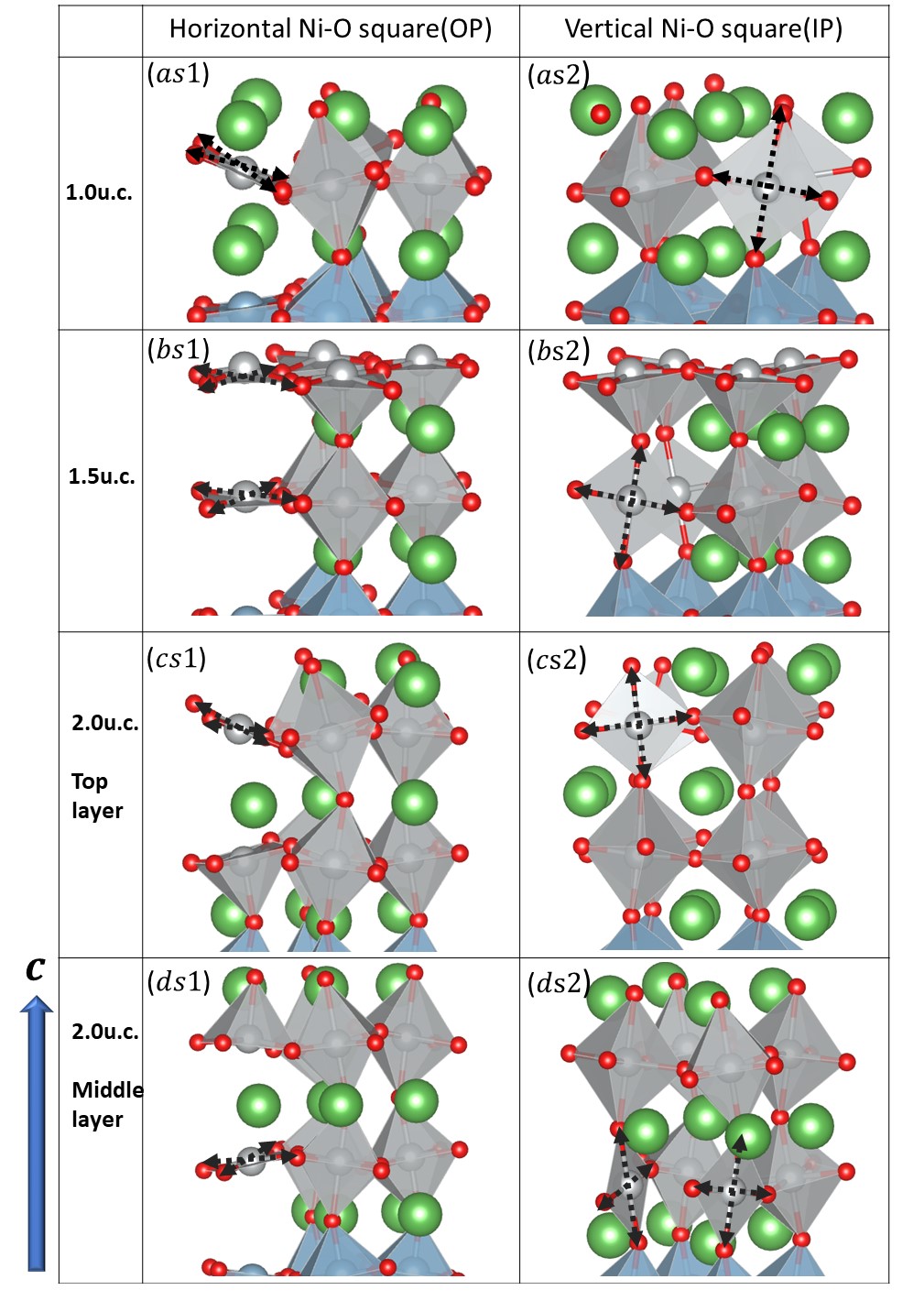}
\caption{
Crystal structure of oxygen defective LaNiO$_3$ thin films with the 0.5-2.0 u.c. thickness. 
Two oxygen-vacant sites are considered for each crystal structure, possibly at OP ($[a-d]s1$) or IP ($[a-d]s2$) sites.
$s$ in the index means that a supercell is constructed to accommodate two vacant sites. 
Black arrow indicates the Ni-O square formation.}
\label{fig:Ovac_sc}
\end{figure}

For thin films thicker than the 0.5 u.c. thickness, we also consider two oxygen vacancies near the Ni site to create a Ni-O square-plane, which is the energetically stable shape in the bulk system with vacancies~\cite{DFT_study}. 
For a certain thickness of thin film, it is important to keep the oxygen vacancy density consistent in different configurations so we can compare formation energies.
To keep the vacancy concentration same as the previous case, we build a $\sqrt{2}\times\sqrt{2}\times1$ supercell containing four Ni ions per layer and remove two oxygen ions.
In this case, we consider both the horizontal Ni-O square (with OP oxygen vacancies; see Fig.\ref{fig:Ovac_sc}$[a-d]s1$) and the vertical Ni-O square (with IP oxygen vacancies; see Fig.\ref{fig:Ovac_sc}$[a-d]s2$). 
For the 2.0 u.c. system there are two Ni layers and we consider the possibility that oxygen vacancies can exist at either layer.

\subsection{\label{sec:FormEnergy}Oxygen vacancy formation energy}
\begin{table*}[ht]
    \centering
    \caption{\label{tbl:FormE_1st2nd}The total energy per 4 Ni ions ($E_{OV}$) computed using DFT+U for the 0.5-2.0 u.c. thin film structures with various oxygen vacancy configurations. Vacancy configuration indices are displayed in the parenthesis. The lowest vacancy configuration energy at each layer is subtracted from $E_{OV}$. } 
    \begin{tabular}{p{0.22\linewidth}p{0.03\linewidth}p{0.14\linewidth}p{0.14\linewidth}p{0.14\linewidth}p{0.12\linewidth}}
    \hline\hline
       &&0.5u.c. & 1.0u.c.&1.5u.c.&2.0u.c.\\ [1.0ex] 
    \hline
       O$_{vac}$ on NiO$_2$ layer   && 0\footnotemark[1](a1)   &0.72eV  (b1)   &0\footnotemark[1](c1) &1.78eV  (d1)\\ 
       O$_{vac}$ on LaO layer       && 0.96eV (a2)      &3.19eV  (b2)   &0.18eV  (c2)   &2.47eV  (d2)\\ 
    \hline
       horizontal Ni-O square       && N/A              &1.83eV (as1)   &2.33eV (bs1)   &0.89eV (cs1)\\
                                    &&&&                                                &0.86eV (ds1)\\
    \hline
       vertical Ni-O square         && N/A              &0\footnotemark[1](as2)&1.83eV (bs2)   &0.72eV (cs2)\\
                                            &&&&                                        &0\footnotemark[1] (ds2)\\
     \hline\hline
    \end{tabular}
    \footnotetext[1]{Lowest energy configuration}
\end{table*} 

In our previous study, we showed that the vacancy formation energy as a function of the oxygen pressure computed using DFT+U is qualitatively similar as the energy curve obtained using more advanced DFT+DMFT calculation while DFT does not show the tendency to form the vacancy due to the underestimation of local correlations~\cite{xingyuliao}.
Here, we calculate the total energies of each vacancy configuration shown in Fig.~\ref{fig:Ovac} and Fig.~\ref{fig:Ovac_sc}. 
For each thickness, we subtract the lowest configuration energy and list in Table \ref{tbl:FormE_1st2nd}.
In the first two rows, we take off one O ion out of 2-Ni-per-layer standard cell, which correspond to Fig.~\ref{fig:Ovac}. In the rest of rows, we take off two O ions from the 4-Ni-per-layer supercell, which are the systems of Fig.~\ref{fig:Ovac_sc}. 
The total energies are normalized to the 4-Ni-per-layer unit cell so that they are comparable each other.
One can see that for configurations shown in Fig.\ref{fig:Ovac}, oxygen vacancies tend to stay on the IP sites rather than the OP sites.
This is because the Ni $d_{x^2-y^2}$ orbital is less occupied than the $d_{z^2}$ orbital and it can accommodate the released charge from the IP vacancy more easily. 
In configurations with two vacancy sites as shown in Fig.\ref{fig:Ovac_sc}, we find that oxygen vacancies stay near the same Ni sites and form Ni-O squares for 1.0u.c. and 2.0u.c. cases. For the 1.5u.c. case, oxygen vacancies still tend to stay on the top-layer IP sites. 
For 1.0u.c. and 2.0u.c. films, vertical square is more favorable than horizontal square, as shown in Fig.\ref{fig:Ovac_sc}-($as2,ds2$). 
This is because the IP vacancy sites are generally favored than the OP sites.

\begin{figure}[b]
       \centering
       \includegraphics[width=1.\linewidth]{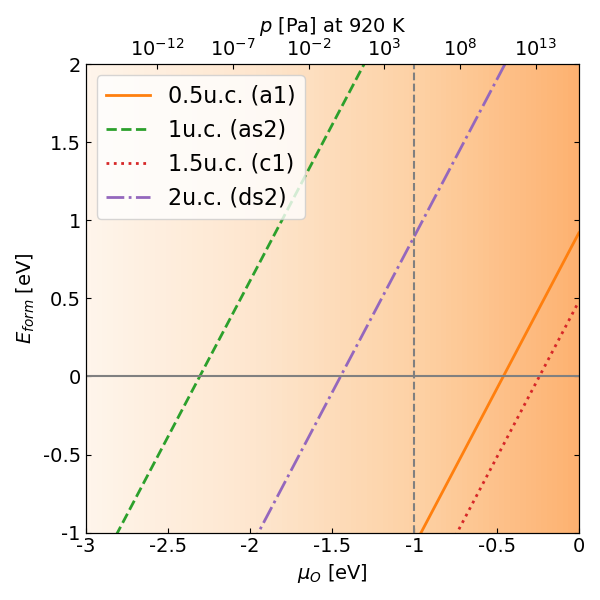}
\caption{The oxygen vacancy formation energy $E_{form}$ as a function of oxygen pressure $p$ at 920K (related to the chemical potential $\mu_O$). 
The negative $E_{form}$ means that the vacancy formation is energetically stable. Dashed line indicates the ambient pressure (10$^5$ Pa).}
\label{fig:FormationE}
\end{figure}

Having found out which oxygen vacancy site is more favorable from Table \ref{tbl:FormE_1st2nd}, we study the stability the oxygen vacancy formation. We consider the formation energy of the oxygen vacancy given by\cite{Geisler}
\begin{equation}
\label{eq:form_E}
E_{form}=E_{OV}-E_{slab}+x\cdot\frac{1}{2}E_{O_2}+x\cdot\mu_O
\end{equation}
where $E_{form}$ is the Gibbs formation energy, and $x$ is the oxygen vacancy concentration. $E_{OV}-E_{slab}$ is the energy difference between systems with and without oxygen vacancy and $\mu_O$ is the oxygen chemical potential.
For each thickness of thin film, we choose the lowest energy configuration for the formation energy calculation, namely (a1) for 0.5u.c., (as2) for 1u.c., (c1) for 1.5u.c. and (ds2) for the 2u.c. system. 
In Fig.~\ref{fig:FormationE}, we plot the formation energy as a function of $\mu_O$. 
The corresponding oxygen pressure is indicated at the top axis.
The relation between oxygen pressure and chemical potential is given by\cite{For1, For2, For3, Geisler}
\begin{equation}
\mu_O(T,P)=\mu_O (T,P^{0})+\frac{1}{2}k_{B}T\ln\left(\frac{P}{P^{0}}\right)
\end{equation}
where $P^{0}$ is the ambient pressure. The $\mu_O (T,P^{0})$ values are measured by Reuter. K et al\cite{For1}.
Positive formation energy system favors state without the oxygen vacancy, and negative formation energy means oxygen vacancy state is more favorable. 
Here, we consider the possible formations of vacancies only at the LaNiO$_3$ layers.
Due to the limit of unit cell size, we consider a fixed oxygen vacancy concentration $x$, namely one vacancy site per two Ni ions.
As shown in Fig.~\ref{fig:FormationE}, both 0.5 and 1.5 u.c. structures have the strong tendency to form vacancies even at the ambient pressure. 
This is because the surface layers of both 0.5 and 1.5 u.c. thin films have the less charge due to the charge reconstruction and they are more adapted to accommodate the released extra charge from oxygen vacancies.
Both 1.0 and 2.0 u.c. structures also can form vacancies at very low oxygen pressures although the 2.0 u.c. structure can form the vacancy more easily compared to the 1.0 u.c. one.

\subsection{\label{sec:vacES}Electronic structure of oxygen defective systems}

\begin{figure}
    \begin{subfigure}{1.0\linewidth}
        \centering
        \includegraphics[width=1.05\linewidth]{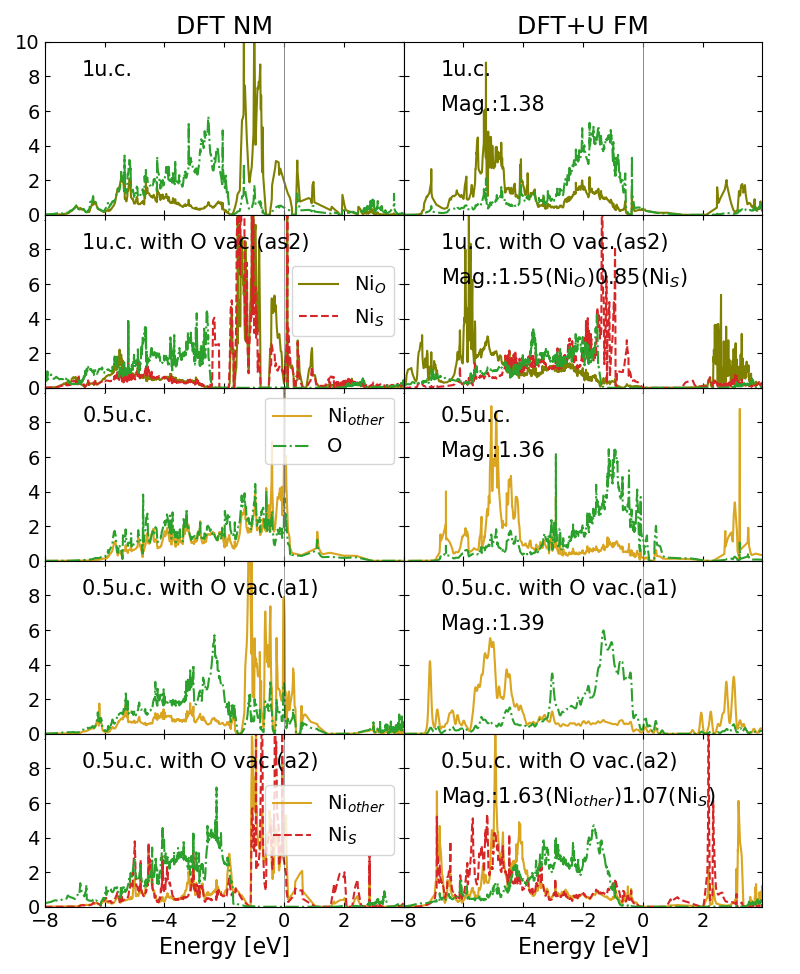}
    \end{subfigure}
    \begin{subfigure}{1\linewidth}
          \centering
       \includegraphics[width=1\linewidth]{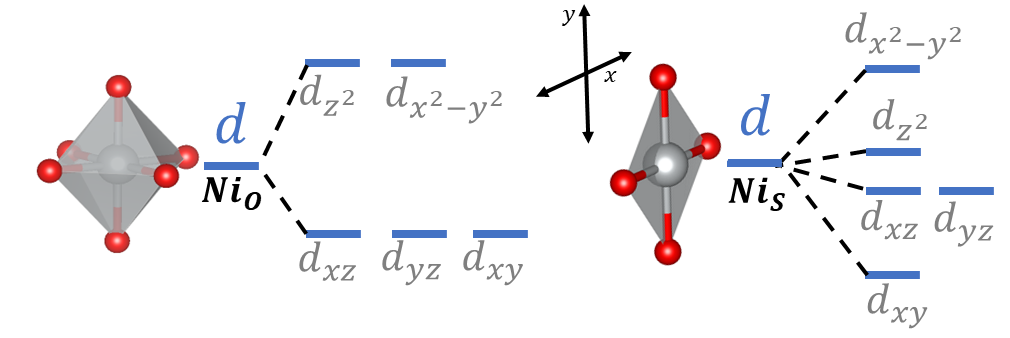}
    \end{subfigure}
\caption{The site and orbital (Ni 3$d$ and O 2$p$) resolved DOS for 0.5 u.c. and 1u.c. structures and those with oxygen vacancies ($a1$, $a2$, and $as2$). 
Both DFT (left panels) and DFT+U (right panels) results are compared.
DFT results are obtained without magnetism and DFT+U calculations are performed with FM (Magnetic moments are included in the figure). 
The energy level diagrams for Ni$_O$ and Ni$_S$ ions are shown in the bottom panel.
The $x-y$ plane of the Ni$_S$ ion is set to be vertical.}
\label{fig:DOS_vac}
\end{figure}

Finally, we discuss how the position of oxygen vacancies can affect the correlated electronic structure of LaNiO$_3$ thin films. 
Here, we focus on the 0.5 and 1.0 u.c. cases to study the different surface termination effect in the ultra-thin layer limit and choose a few oxygen defective structures with the lowest energies.
Fig.~\ref{fig:DOS_vac} shows the site-resolved Ni $3d$ and O $2p$ DOS for the 0.5 and 1.0 u.c. thin films with and without oxygen vacancies.
In DFT (see Fig.~\ref{fig:DOS_vac} left panel), all vacancy structures we considered show the metallic behavior exhibiting the spiky nature of the DOS due to the non-bonding nature of $t_{2g}$ orbitals. Compared to the structure without the oxygen vacancy, the O $p$ DOS has pushed below the Fermi energy as the released electrons from the vacancy are occupied in the O $p$ state. In DFT+U, all oxygen defective structures show the strong reduction of the DOS at the Fermi energy compared to the DFT DOS although the insulating state with the full gap opening can be emerged depending on the vacancy positions. 

On the top two panels of Fig.~\ref{fig:DOS_vac}, we plot the DOS for the LaO-terminated structure (1 u.c.) and its oxygen vacancy structure $as2$ (see Fig.~\ref{fig:Ovac_sc}), which is the lowest energy configuration among the 1 u.c. systems.
In the oxygen defective structure $as2$, two types of Ni environments are created with the distinct crystal field splitting, as shown in the lower panel of Fig.~\ref{fig:DOS_vac}. Ni$_O$, with the octahedral geometry, has $e_g$-$t_{2g}$ energy splitting under the cubic crystal symmetry, while Ni$_S$ has the lower symmetry due to the square-plane geometry and exhibits the further energy splitting as the cubic symmetry is broken within $e_g$ and $t_{2g}$ orbitals. 
In the DFT spectra, the oxygen defective system shows no significant difference between the Ni$_O$ and Ni$_S$ DOS with the metallic behavior at the Fermi energy.
However, the DOS is strongly modified in DFT+U as two Ni ions have the distinct correlation effect in the the oxygen defective structure.
Although the $d$ orbital occupancy of both Ni ions is close to 8.0,  the Ni$_O$ DOS opens a rather large spectral gap driven by the strong correlation of Ni$_O$ orbitals while the Ni$_S$ DOS shows a similar feature as the DFT one with the weaker correlation and the $d_{z^2}$ spectra are pushed below the Fermi energy due to the large splitting between the $d_{z^2}$ and $d_{x^2-y^2}$ orbitals.
The spectral gap size is as large as $\sim$1eV, which is consistent with the gap size measured in the experimental PES of the LaNiO$_3$ thin film below 2.0 u.c.~\cite{doi:10.1063/1.4916225}.  
This different degrees of correlations are also reflected in the magnetic moments since the Ni$_O$ ion has the large moment of 1.55$\mu_B$ but the Ni$_S$ ion shows the smaller moment of 0.85$\mu_B$ due to the weaker correlation compared to the moment of 1.38$\mu_B$ measured in the defect-free 1.0 u.c. structure showing the pseudo-gap feature at the Fermi energy.
This novel electronic behavior with two distinct correlation effects is also reminiscent of the ``site-selective'' Mott phase found in the oxygen defective LaNiO$_{2.5}$ with two nonequivalent Ni ions~\cite{xingyuliao}.

In the 0.5 u.c. and its oxygen defective structure ($a1$), the Ni ions have other environments than the octahedron (Ni$_O$) and the square-plane (Ni$_S$) mentioned above. Here, we note the average spectra of all other Ni ions except the Ni$_O$ and Ni$_S$ as Ni$_{other}$. 
Compared to the defect-free 0.5 u.c. case, the lowest-energy defective $a1$ structure shows the similarity in the DOS displaying the pseudo-gap feature at the Fermi energy with the similar magnetic moments of 1.39$\mu_B$.  
To study the dependence of the vacancy position on the MIT property, we also study the oxygen defective $a2$ structure using DFT+U.
The $a2$ structure has the Ni-O square-plane geometry similarly as the $a1$ structure, as well as the pyramid geometry.
Remarkably, the DFT+U DOS of the $a2$ structure shows the insulating behavior with a gap slightly smaller than 1eV.
Similarly as the $as2$ structure, the Ni$_S$ ion shows the smallest moment of 1.07$\mu_B$ while the other Ni ions have the enhanced moment of 1.63$\mu_B$ due to the strong correlation effect.
Therefore, the formation of the square-planar Ni ion exhibits an insulating behavior driven by the large energy splitting between $d_{z^2}$ and $d_{x^2-y^2}$ orbitals with the weak correlation effect, and at the same time, it promotes a Mott insulating state for other Ni ions with large magnetic moments.
Our result shows that the MIT behavior in LaNiO$_3$ thin films can be tuned by different oxygen vacancy positions, which induce the distinct correlation strength of Ni ions due to different chemical environments under the presence of vacancies.

\section{\label{sec:summary}Summary}
In summary, we perform first-principles DFT+U calculations to study the electronic structure of LaNiO$_3$ ultra-thin films on the compressive LaAlO$_3$ substrate and the effect of oxygen vacancies on the MIT systematically.
In the LaNiO$_3$ thin films without explicit oxygen vacancies, the electronic structure is strongly modified from bulk due to different surface terminations, the strain effect, the quantum confinement effect, and the charge reconstruction.
Different surface terminations of the polar layer in thin film dictate the charge reconstruction of the surface.
The NiO$_2$ (LaO) terminated structure has the less (more) charge density on the surface layer compared to the bulk one and, as a result, the relative positions of Ni and O spectra are strongly modified depending on the surface termination.
Although the Ni $d$ orbital occupancy is close to 8.0 and it does not vary much across the layers, the O $p$ DOS is mostly affected and modified compared to the bulk case.
The Ni $d_{z^2}$ orbital is also strongly renormalized and more occupied on the surface layer of the NiO$_2$ terminated structures.
The DFT+U DOS exhibits a pseudo-gap state with the strong reduction of the DOS at the Fermi energy in all structures although the surface termination and the orbital anisotropy effects become weaker compared to the DFT DOS.

When the oxygen vacancies are present in LaNiO$_3$ thin films, they are energetically favored to stay in the NiO$_2$ layers rather than in the LaO layers, and the vacancy formation energy is substantially lower for the NiO$_2$ terminated structures (0.5 and 1.5 u.c.) compared to the LaO terminated structures.
Since the NiO$_2$ terminated structures have less charges compared to bulk due to the charge reconstruction, the charge release due to the vacancy formation can be favored.
In both 1.0 and 2.0 u.c. structures, two vacancies tend to stay close to a Ni ion and form a square-plane Ni-O geometry.
The presence of oxygen vacancies also strongly affects the MIT in the thin film.
In both 0.5 u.c. and 1.0 u.c. cases, the oxygen defective structure containing the Ni-O square-plane becomes an insulating phase with a sizable gap size of $\sim$1eV within DFT+U.
This is because the square-plane Ni ion shows the large energy splitting between $d_{z^2}$ and $d_{x^2-y^2}$ orbitals and the insulating state develops with a rather small magnetic moment while other Ni ions undergo a Mott transition of an insulating phase with a large magnetic moment.
Therefore, it is important to treat nonequivalent Ni ions under different chemical environments independently in the realistic oxygen defective structures as they show distinct correlation strengths and modify the electronic structure significantly.
Our results suggest that the oxygen vacancy position can strongly affect the MIT occurring in transition metal oxide thin films and the first-principles description going beyond the rigid-band shift approximation will be crucial to capture such novel electronic behavior.   

\section*{Acknowledgement}
This work is supported by ACS-PRF grant 60617. H. Park is also supported by the Materials Sciences and Engineering Division, Basic Energy Sciences, Office of Science, US DOE. We gratefully acknowledge the computing resources provided on Bebop, a high-performance computing cluster operated by the Laboratory Computing Resource Center at Argonne National Laboratory.

\bibliography{ref.bib}

\end{document}